\documentclass[prb,showpacs,twocolumn,nofloats,endfloats]{revtex4}

\usepackage{amsmath,amssymb,epsfig,epsf}

\begin{document}

\title
{Pseudogap behaviour in Bi$_2$Ca$_2$SrCu$_2$O$_8$: Results of Generalized Dynamical Mean-Field Approach}
\author{E.Z.Kuchinskii$^1$, I.A.Nekrasov$^{1}$, Z.V.Pchelkina$^2$,M.V.Sadovskii$^1$}

\affiliation
{$^1$Institute for Electrophysics, Russian Academy of Sciences,
Ekaterinburg, 620016, Russia\\
$^2$Institute for Metal Physics, Russian Academy of Sciences, 
Ekaterinburg, 620219, Russia} 

\begin{abstract}


Pseudogap phenomena are observed for normal underdoped phase of different 
high-T$_c$ cuprates.
Among others Bi$_2$Sr$_2$CaCu$_2$O$_{8-\delta}$ (Bi2212) compound is one 
of the most studied experimentally.
To describe pseudogap regime in Bi2212 we employ novel generalized $ab~initio$
LDA+DMFT+$\Sigma_{\bf k}$ hybrid scheme.
This scheme based on the strategy of one of the most powerfull computational tool for real 
correlated materials:
local density approximation (LDA) + dynamical mean-field theory (DMFT).
Here conventional LDA+DMFT equations are supplied by an additional (momentum 
dependent) self-energy $\Sigma_{\bf k}$ in the spirit of our recently 
proposed DMFT+$\Sigma_{\bf k}$ approach, accounting for pseudogap fluctuations.  
In the present model $\Sigma_{\bf k}$ describes non-local 
correlations induced by short-ranged collective Heisenberg-like 
antiferromagnetic spin fluctuations. The effective single impurity problem of 
the DMFT is solved by numerical renormalization group (NRG). Material 
specific model parameters for effective $x^2-y^2$ orbital of Cu-3d shell of Bi2212 compound,
e.g. the values of intra- and interlayer hopping 
integrals between different Cu sites, local Coulomb interaction U
and pseudogap potential $\Delta$ were obtained within LDA and LDA+DMFT.  
Here we report theoretical LDA+DMFT+$\Sigma_{\bf k}$ quasiparticle bands 
dispersion and damping, Fermi surface renormalization, momentum anisotropy of 
(quasi) static scattering, densities of states, spectral densities and angular resolved 
photoemission (ARPES) spectra accounting for pseudogap and bilayer splitting 
effects for normal (slightly) underdoped Bi2212 ($\delta$=0.15). We show that 
LDA+DMFT+$\Sigma_{\bf k}$ successfully describes strong (pseudogap) 
scattering close to Brillouin zone boundaries. Our calculated 
LDA+DMFT+$\Sigma_{\bf k}$ Fermi surfaces and ARPES spectra in presence of the pseudogap 
fluctuations are almost insensitive to the bilayer splitting strength.  
However, our LDA-calculated value of bilayer splitting is found to be rather small to describe 
experimentally observed peak-dip-hump structure.  Results obtained are in 
good semiquantitative agreement with various recent ARPES experiments.  

\normalsize

\end{abstract}


\maketitle

\newpage

\section{Introduction}

Pseudogap state is the major anomaly of the normal state of copper oxides, 
commonly believed to be most relevant for the understanding of the physical 
nature of high-$T_c$ superconductivity \cite{psgap}.

During the last decade experimental techniques of angular resolved photoemission 
spectroscopy (ARPES) has made a large progress. Sate-of-the-art high-T$_c$ 
test compound for ARPES is Bi$_2$Sr$_2$CaCu$_2$O$_{8-\delta}$ 
(Bi2212) system. Thus a lot of experimental ARPES data is available for Bi2212 
(for reviews see \cite{Bi2212}). Several major experimental characteristics
are derived from ARPES data like, for example, Fermi surfaces (FS), 
quasiparticle band dispersions and damping, and even self-energy lineshapes  
\cite{Bi2212}. A number of interesting physical anomalies were discovered in 
normal underdoped phase of Bi2212: pseudogap formation, shadow bands and 
bilayer splitting of FS \cite{Bi2212}. These phenomena abounds in theories 
and there are still no definite point of view about their physical origin. 
It is believed that all of these are quite relevant for the physics of
high-temperature superconductors. The problem is much complicated by strong
electronic correlations ever present in these compounds, making the standard
band theory and Fermi liquid approaches doubtful.

In this work we show that taking into account short-range antiferromagnetic 
fluctuations resulting in pseudogap formation together with bilayer splitting 
effects is enough, in principle, to describe abovementioned experiments. 
To this end we employ novel hybrid $ab~initio$ LDA+DMFT+$\Sigma_{\bf k}$
computational scheme \cite{jtl,cm05,FNT}. 
From one side this scheme inherits all the advantages
of LDA+DMFT \cite{poter97,LDADMFT1,Nekrasov00, psik, LDADMFT}, i.e.
the merger of first principle one-electron density functional theory within
local densty approximation (DFT/LDA) \cite{DFT_LDA,DFT_LDA_1} and the dynamical 
mean-field theory (DMFT) for strongly correlated
electrons \cite{MetzVoll89,vollha93,pruschke,georges96,PT}.
On the other side our scheme allows one to account for non-local correlation
effects introducing momentum dependent external self-energy preserving 
conventional DMFT equations \cite{jtl,cm05,FNT}.
To solve the effective single impurity problem of the DMFT we
employ here the reliable numerical renormalization group approach (NRG) 
\cite{NRG,BPH}.

Such combined scheme is particularly suitable to describe electronic 
properties of real high-T$_c$ materials at finite doping in the normal state.
First, all material specific model parameters
for physically relevant effective $x^2-y^2$ orbital of Cu-3d shell are obtained from LDA 
computations. Second, undoped cuprates are antiferromagnetic Mott insulators with
$U\gg W$ ($U$ --- value of local Coulomb interaction, $W$ --- bandwidth of
non--interacting band), so that correlation effects are very important.
Thus at finite doping (up to optimal doping) cuprates are typical strongly 
correlated metals. To this end DMFT stage in our computational scheme takes these 
strong electronic correlations into account. To adopt LDA+DMFT to study the 
``antiferromagnetic'' scenario of pseudogap formation in 
cuprates \cite{psgap,Pines,Sch,KS} {\bf k}-dependent self-energy 
$\Sigma_{\bf k}$ describing non-local correlations induced by 
(quasi) static short-ranged collective Heisenberg-like antiferromagnetic (AFM)
spin fluctuations is included \cite{Sch,KS}.

Recently we applied DMFT+$\Sigma_{\bf k}$ approach to
investigate formation of pseudogap for strongly correlated metallic regime of 
single-band Hubbard model on the square lattice \cite{jtl,cm05,FNT}.
At present there are several independent methods aimed to describe 
non-local effects beyond standard DMFT.  Similar results about pseudogap 
formation in the 2d Hubbard model were already obtained within 
two-particle self-consistent approach \cite{Kyung06}, cluster DMFT 
extensions, such as the dynamical cluster approximation (DCA) 
\cite{TMrmp,mpj03} and cellular DMFT (CDMFT) \cite{KSPB,Kyung05,Kyung06,CivKot}, 
CPT \cite{Gross94,Senechal00,Senechal05} and via the model of two interacting
Hubbard sites self-consistently embedded in a bath \cite{Stanescu03}.
The EDMFT was employed to demonstrate
pseudogap formation in the DOS due to dynamic Coulomb correlations \cite{HW}.
Important progress was also made  with weak coupling approaches for the
Hubbard model \cite{KyTr} and functional renormalization group 
\cite{Katanin04,RM}. In several papers pseudogap formation was described in 
the framework of the t-J model \cite{Prelovsek}.  A more general scheme for 
the inclusion of non-local corrections was also formulated within the so 
called GW extension to DMFT \cite{BAG,SKot}.  Dynamical vertex 
approximation to study Mott-Hubbard transition in presence of non-local 
antiferromagnetic correlations was proposed \cite{Held05}. Chain-DMFT 
extension was used to investigate breakup of the Fermi surface near Mott 
transition for quasi 1d Hubbard model \cite{Georges06}.

The paper is organized as follows.
In section \ref{method} we present a short introduction into an
$ab~initio$ self-consistent generalized combined LDA+DMFT+$\Sigma_{\bf k}$ 
scheme to account for short-rang AFM correlations.
Section \ref{lda} contains Bi2212 material specific information:\ 
LDA calculated band structure and details on some model parameters 
calculations.  Results and a discussion of LDA+DMFT+$\Sigma_{\bf k}$ 
computations for Bi2212 are presented in the sections \ref{results} and 
\ref{concl}.

\section{Computational method}
\label{method}

\subsection{Introduction of the length scale into DMFT: DMFT+$\Sigma_{\bf k}$ 
approach} \label{leng_intro}

To introduce spatial length scale into conventional
DMFT method \cite{MetzVoll89,vollha93,pruschke,georges96,PT}
recently we proposed generalized DMFT+$\Sigma_{\bf k}$ approach \cite{jtl,cm05,FNT}.
The major assumption of our approach is that the lattice
and Matsubara ``time'' Fourier transform of the single-particle Green function
can be written as
\begin{equation}
G(\omega,{\bf k})=\frac{1}{i\omega+\mu-\varepsilon({\bf k})-\Sigma(\omega)
-\Sigma_{\bf k}(\omega)},
\label{Gk}
\end{equation}
where $\Sigma(\omega)$ is the {\em local} self--energy of DMFT,
while $\Sigma_{\bf k}(\omega)$ is some momentum dependent part.
Interference effects between these parts are neglected.
Advantage of our generalized DMFT+$\Sigma_{\bf k}$ approach is additive form of
self-energy in Eq. (\ref{Gk}) \cite{jtl,cm05,FNT}.
It allows one to keep the set of self-consistent equations of standart DMFT
\cite{MetzVoll89,vollha93,pruschke,georges96,PT}.
However there are two distinctions. First, on each DMFT iteration we recalculate
corresponding  {\bf {k}}-dependent self-energy
$\Sigma_{\bf k}(\mu,\omega,[\Sigma(\omega)])$ within some (approximate) scheme,
taking into account interactions with collective modes or order parameter
fluctuations.
Second, the local Green's function of effective impurity problem is defined as
\begin{equation}
G_{ii}(\omega)=\frac{1}{N}\sum_{\bf k}\frac{1}{i\omega+\mu
-\varepsilon({\bf k})-\Sigma(\omega)-\Sigma_{\bf k}(\omega)},
\label{Gloc}
\end{equation}
at each step of the standard DMFT procedure.

Eventually, we get the desired Green function in the form of (\ref{Gk}),
where $\Sigma(\omega)$ and $\Sigma_{\bf k}(\omega)$ are those appearing
at the end of our iteration procedure.

To calculate $\Sigma_{\bf k}(\omega)$ for an electron moving in the
random field of pseudogap fluctuations (assumed to be (quasi) static and 
Gaussian, which is valid at high enough temperatures \cite{Sch,KS}) with 
dominant scattering momentum transfers of the order of characteristic vector 
${\bf Q}=(\pi/a,\pi/a)$ ($a$ - lattice spacing) of AFM fluctuations (``hot 
spots'' model \cite{psgap}), we use the following recursion procedure 
proposed in Refs.~\cite{Sch,KS,MS79}
\begin{equation} 
\Sigma_{\bf k}(\omega)=\Sigma_{n=1}(\omega,{\bf k}), \label{Sk} 
\end{equation} 
where 
\begin{equation} 
\Sigma_{n}(\omega,{\bf k})= 
\Delta^2\frac{s(n)}{i\omega+\mu-\Sigma(\omega) -\varepsilon_n({\bf k})
+inv_n\kappa-\Sigma_{n+1}(\omega,{\bf k})}.  
\label{rec} 
\end{equation} 
\normalsize
The quantity $\Delta$ characterizes the pseudogap energy scale and
$\kappa=\xi^{-1}$ is the inverse correlation length of short range
SDW fluctuations, $\varepsilon_n({\bf k})=\varepsilon({\bf k+Q})$ and
$v_n=|v_{\bf k+Q}^{x}|+|v_{\bf k+Q}^{y}|$
for odd $n$, while $\varepsilon_n({\bf k})=\varepsilon({\bf k})$ and $v_{n}=
|v_{\bf k}^x|+|v_{\bf k}^{y}|$ for even $n$ with $v^{x,y}({\bf p})$
determined by usual momentum derivatives of the ``bare'' dispersion
$\varepsilon({\bf k})$, while $s(n)$ represents a combinatorial factor,
determining the number of Feynman diagrams \cite{Sch,KS}.

For the (Heisenberg) spin structure of interaction with 
spin fluctuations in  ``nearly antiferromagnetic Fermi--liquid'' 
(spin--fermion (SF) model of Ref.~\cite{Sch})
spin-conserving scattering processes obey commensurate combinatorics,
while spin--flip scattering is described by diagrams of incommensurate
type (``charged'' random field in terms of Ref.~\cite{Sch}). In this model
combinatorial factor $s(n)$ acquires the following form \cite{Sch}
\begin{equation} 
s(n)=\left\{\begin{array}{cc}
\frac{n+2}{3} & \mbox{for odd $n$} \\
\frac{n}{3} & \mbox{for even $n$}.
\end{array} \right.
\label{vspin}
\end{equation}

Obviously, with this procedure we introduce an important length scale $\xi$ 
not present in conventional DMFT. Physically this scale mimics the effect of 
short-range (SDW) fluctuations within fermionic ``bath'' surrounding 
the effective Anderson impurity of the DMFT. We expect that such a 
length-scale dependence will lead to a kind of competition between local and 
non-local physics.

Though we prefer to consider both parameters $\Delta$ and $\xi$ as 
phenomenological (to be determined by fitting experiments) \cite{cm05}, one
can in principle calculate these from microscopic model at hand. For 
example, using the two-particle self-consistent approach of 
Refs.~\cite{Kyung06,VT} with the approximations introduced in 
Refs.~\cite{Sch,KS}, we derived 
within the standard Hubbard model the following microscopic 
expression for $\Delta$ \cite{cm05}
\begin{equation} 
\Delta^2=U^2\frac{<n_{i\uparrow}n_{i\downarrow}>}{n^2}<(n_{i\uparrow}
-n_{i\downarrow})^2>,
\label{DeltHubb}
\end{equation}
where we consider only scattering by antiferromagnetic spin fluctuations.
The different local quantities -- total density $n$,
local densities $n_{i\uparrow}$, $n_{i\downarrow}$ and  double occupancy
$<n_{i\uparrow}n_{i\downarrow}>$ -- can easily be calculated within the 
standard DMFT \cite{georges96}. A detailed derivation of (\ref{DeltHubb}) 
is presented in the Appendix B of Ref.~\cite{cm05}.
Corresponding microscopic expressions for the correlation length 
$\xi=\kappa^{-1}$ can also be derived within the two--particle 
self--consistent approach \cite{Kyung06,VT}. However, we expect these results 
for $\xi$ to be less reliable, because this approach is valid only for 
relatively small (or medium) values of $U/t$, as well as for purely 
two-dimensional case, neglecting  quasi-two-dimensional effects, obviously
important for cuprates. Actually, our calculation experience shows that all 
the results obtained below are rather weakly dependent on the values of
$\xi$ from the experimentally relevant \cite{psgap} interval of $(5\div 10)a$.

\subsection{Bilayer splitting effects: 
LDA+DMFT+$\Sigma_{\bf k}$ formulation}

To perform $ab~initio$ calculations for Bi2212 system we employ LDA+DMFT strategy
proposed in Refs.~\cite{poter97,LDADMFT1,Nekrasov00, psik, LDADMFT}.
Necessary bare band dispersion for effective physically relevant Cu-3d $x^2-y^2$ orbital
in a tight-binding representation is
\begin{eqnarray}\label{disp}
\varepsilon({\bf k})=&-2t&(\cos k_xa+\cos k_ya)\\ \nonumber
&-4t'&\cos k_xa\cos k_ya\\ \nonumber
&-2t''&(\cos 2k_xa+\cos 2k_ya)\\ \nonumber
&-2t'''&(\cos k_xa\cos 2k_ya +\cos 2k_ya \cos k_ya).
\end{eqnarray}
Here $t$, $t^{\prime}$, $t^{\prime\prime}$, $t^{\prime\prime\prime}$ are hopping
integrals within first four coordination spheres. Tight-binding equation
for interlayer dispersion is taken in the form
\begin{equation}
t_{\perp}({\bf k}) = \frac{t_{\perp}}{4}(\cos k_xa-\cos k_ya)^2
\label{tperp}
\end{equation}
given in Ref.~\cite{Andersen95} with bilayer splitting equal to 2$t_\perp$.

Since account of bilayer splitting (BS) effects in Bi2212 requires essentially 
two-band model we introduce bare Hamiltonian in reciprocal space as the
following matrix over (bonding and antibonding) band indices
\begin{equation}
{\bf \hat H}({\bf k}) = \left( \begin{array}{cc} \varepsilon({\bf k}) &
t_{\perp}({\bf k}) \\
t_{\perp}({\bf k})  & \varepsilon({\bf k}) \end{array} \right).
\label{bareham}
\end{equation}
The local Green's function is now also a matrix
\begin{equation}
{\bf\hat G}(\omega)=\frac{1}{N}\sum_{\bf k}
\biggl(i\omega-{\bf \hat H}({\bf k})-(\Sigma(\omega)+
\Sigma_{\bf k}(\omega)){\bf\hat I}\biggr)^{-1}.
\label{latprob}
\end{equation}
where we have assumed self-energies to be diagonal.
In the following we would like to keep the DMFT part of the problem 
just a single-band task. This can be achieved taking the diagonal element of
(\ref{latprob})
\begin{equation}
{\widetilde G}(\omega) = \frac{1}{N}\sum_{\bf k} \frac{G^{-1}(\omega,{\bf k})}
{(G^{-1}(\omega,{\bf k}))^2-(t_\perp({\bf k}))^2},
\label{diagG}
\end{equation}
where $G(\omega,{\bf k})$ is given by (\ref{Gk}).
This local Green's function ${\widetilde G}(\omega)$ (which includes 
additive self-energy contributions) determines now our effective single 
Anderson impurity problem. One should remark since we work
with the single-band problem there is no need for double counting
correction between LDA and DMFT.\cite{Nekrasov00}

\section{LDA band structure of Bi2212 and effective model parameters}
\label{lda}

The Bi2212 compound has tetragonal bcc crystal lattice with symmetry space 
group I4/mmm  \cite{Hybersten88,Tarascon88,Sunshine88}. Main structural motif 
for this compound is two CuO$_2$
layers displaced close to each other in the unit cell. 
Using crystal structure data of
Ref.~\cite{Hybersten88} we performed LDA calculations
of electronic band structure within the linearized muffin-tin orbital (LMTO) 
basis set \cite{LMTO}.
Obtained band structure is in agreement with the one of Ref. 
\cite{Hybersten88}. 

\begin{figure}[htb]
\includegraphics[clip=true,width=0.4\textwidth]{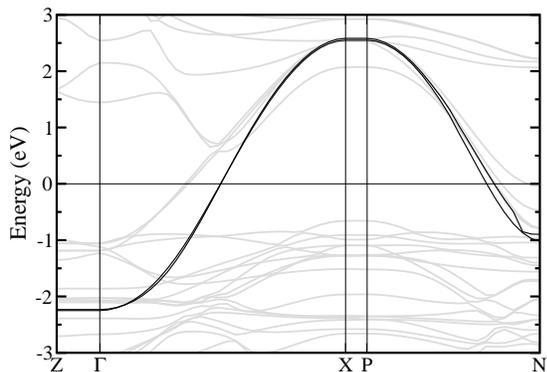}
\caption{Calculated within DFT/LDA Bi2212 band dispersions (gray lines) and
effective $x^2-y^2$ band of Cu-3d shell obtained by projection on Wannier functions (black lines).
The Fermi level corresponds to zero.}
\label{ldabands}
\end{figure}

In Fig.~\ref{ldabands} one-electron LDA band dispersion along BZ symmetry lines 
for Bi2212 is shown. Gray lines correspond
to all-band Hamiltonian. To extract physically interesting partially filled 
$x^2-y^2$ orbital of Cu-3d shell
Wannier functions projecting method \cite{Marzari} in the LMTO framework 
\cite{Anisimov05} was applied.
Corresponding dispersion of effective $x^2-y^2$ orbital is displayed in 
Fig.~\ref{ldabands} as a black line.

To set up the LDA+DMFT+$\Sigma_{\bf k}$ lattice problem (\ref{Gloc}) 
one needs to calculate transfer integrals
$t$, $t^{\prime}$, $t^{\prime\prime}$, $t^{\prime\prime\prime}$ and 
$t_\perp$ for tight-binding expressions (\ref{disp}) and (\ref{tperp}). On the 
basis of Wannier function projecting \cite{Marzari} we performed computation of 
corresponding hopping integrals with its LMTO realization \cite{Anisimov05}. 
Obtained values for intra- and interlayer hybridization between $x^2-y^2$ 
orbital of different Cu-sites are listed in the Table I.  Values of 
$t$, $t^{\prime}$, $t^{\prime\prime}$, $t^{\prime\prime\prime}$ we present 
are somewhat larger than those extracted from ARPES 
experiment \cite{Kordyuk03}. On the 
other hand our value of $t_\perp$ is much smaller than experimental one 
$t^{exp}_\perp$=0.083~eV \cite{Kordyuk03}. At the same time our calculated 
value of $t_\perp$ is in good agreement with other band structure results 
reported \cite{Kordyuk04}. Taking into account large difference between 
$t_\perp$ and $t^{exp}_\perp$ further we provide LDA+DMFT+$\Sigma_{\bf k}$ 
results for both these values.  

The value of local Coulomb interaction U for $x^2-y^2$ orbital was obtained  
via constrained LDA method \cite{Gunnarsson88}.
To screen this $x^2-y^2$ orbital we used the rest of the Cu-3d shell of our 
selected site, neighbouring inplane Cu sites and also 
Cu sites from closest CuO$_2$ layer. The value found is U=1.51 eV (Table I).

Pseudogap potential $\Delta$ (see Eq. (\ref{DeltHubb})) was obtained as 
described in Ref.~\cite{cm05} using LDA+DMFT(NRG) to calculate set of occupancies entering
(\ref{DeltHubb}) (instead of DMFT(QMC) used in Ref. \cite{cm05}). 
For given values of hopping integrals and U value with hole doping level 
$\delta=0.15$ our $\Delta$ equals 0.21~eV.  The value of correlation length 
$\xi$ is always taken to be equal to 5 lattice constants which is a typical 
experimental value~\cite{psgap}.
Temperature comes through NRG part of our scheme and is always taken to be  
$\sim$255~K.  This completes the set of 
necessary model parameters to start LDA+DMFT+$\Sigma_{\bf k}$ 
computations for Bi2212 (see Sec.~\ref{method}).

\section{Results and discussion}
\label{results}

\subsection{Bi2212 LDA+DMFT+$\Sigma_{\bf k}$ densities of states}

\begin{figure}[htb]
\includegraphics[clip=true,width=0.4\textwidth]{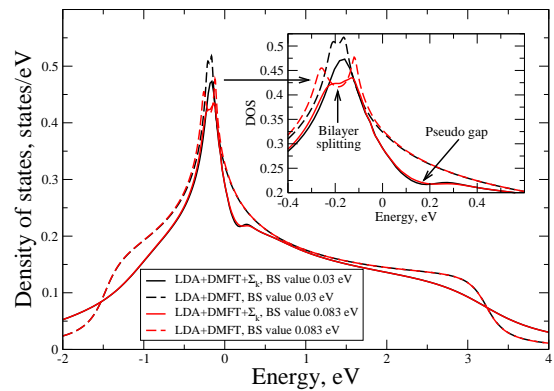}
\caption{LDA+DMFT (dashed lines) and LDA+DMFT+$\Sigma_{\bf k}$ (solid lines)
densities of states for Bi2212
for LDA-calculated value of $t_\perp$=0.03 eV (black) and experimental value of $t^{exp}_\perp$=0.083 eV (gray)
(Coulomb interaction U=1.51 eV, filling n=0.85, pseudogap potential $\Delta$=0.21 eV, correlation length $\xi=5a$).
Inset shows magnified region around the Fermi level.}
\label{dos}
\end{figure}

Density of states (DOS) is calculated as 
\begin{equation}
N(\omega)=-\frac{1}{\pi}{\rm Im}\widetilde G(\omega)
\label{DOS}
\end{equation}
where $\widetilde G(\omega)$ is defined by Eq.~\ref{diagG} analytically continued
to real frequencies.
In Fig.~\ref{dos} we display LDA+DMFT and LDA+DMFT+$\Sigma_{\bf k}$ DOS 
for effective $x^2-y^2$ orbital of Cu-3d. It is clearly seen that pseudogap
fluctuations lead to formation of the pseudogap in DOS within 0.2 eV from the
Fermi level. In our model this pseudogap is not tied to the Fermi level and it
is not very pronounced for parameter values used here for Bi2212.
It is also easy to find out that for all our DOS curves BS effects are 
most pronounced on the top of the van Hove singularity, which is about
around -0.2 eV below the Fermi level (see also inset of Fig.~\ref{dos} for 
details). Namely we calculate DOS for LDA value of BS 0.03 eV (gray curve) and 
experimental BS value of 0.083 eV (black line). For the latter case BS effects 
are obviously stronger.
Dashed curves correspond to LDA+DMFT results for two different values of bilayer 
splitting.
For LDA+DMFT+$\Sigma_{\bf k}$ DOS (solid curves) it is observed that BS effects 
become less pronounced (but still can be seen for the case of 
$t^{exp}_\perp$=0.083 eV).  
This is caused by the decrease of the life-time due to pseudogap fluctuations. 
Also van Hove singularity becomes slightly narrower here due to self-energy 
effects. Note that the shape of the pseudo gap in the DOS almost does not depend 
on BS effects.

\subsection{Bi2212 LDA+DMFT+$\Sigma_{\bf k}$ qusiparticle dispersions
and damping}

For the case of finite temperature and interaction values we define 
quasiparticle dispersions via maxima positions of corresponding spectral 
functions $A(\omega,{\bf k})$
\begin{equation} 
A(\omega,{\bf k})=-\frac{1}{\pi}{\rm Im}\widetilde G(\omega,{\bf k}), 
\label{specf} 
\end{equation} 
where $\widetilde G(\omega,{\bf k})$ is defined by an expression under the sum
in (\ref{diagG}), analytically continued to real frequencies, with
self--energies and chemical potential $\mu$
calculated self--consistently as described in Sec. \ref{leng_intro}.

\begin{figure}[htb]
\includegraphics[clip=true,width=0.4\textwidth]{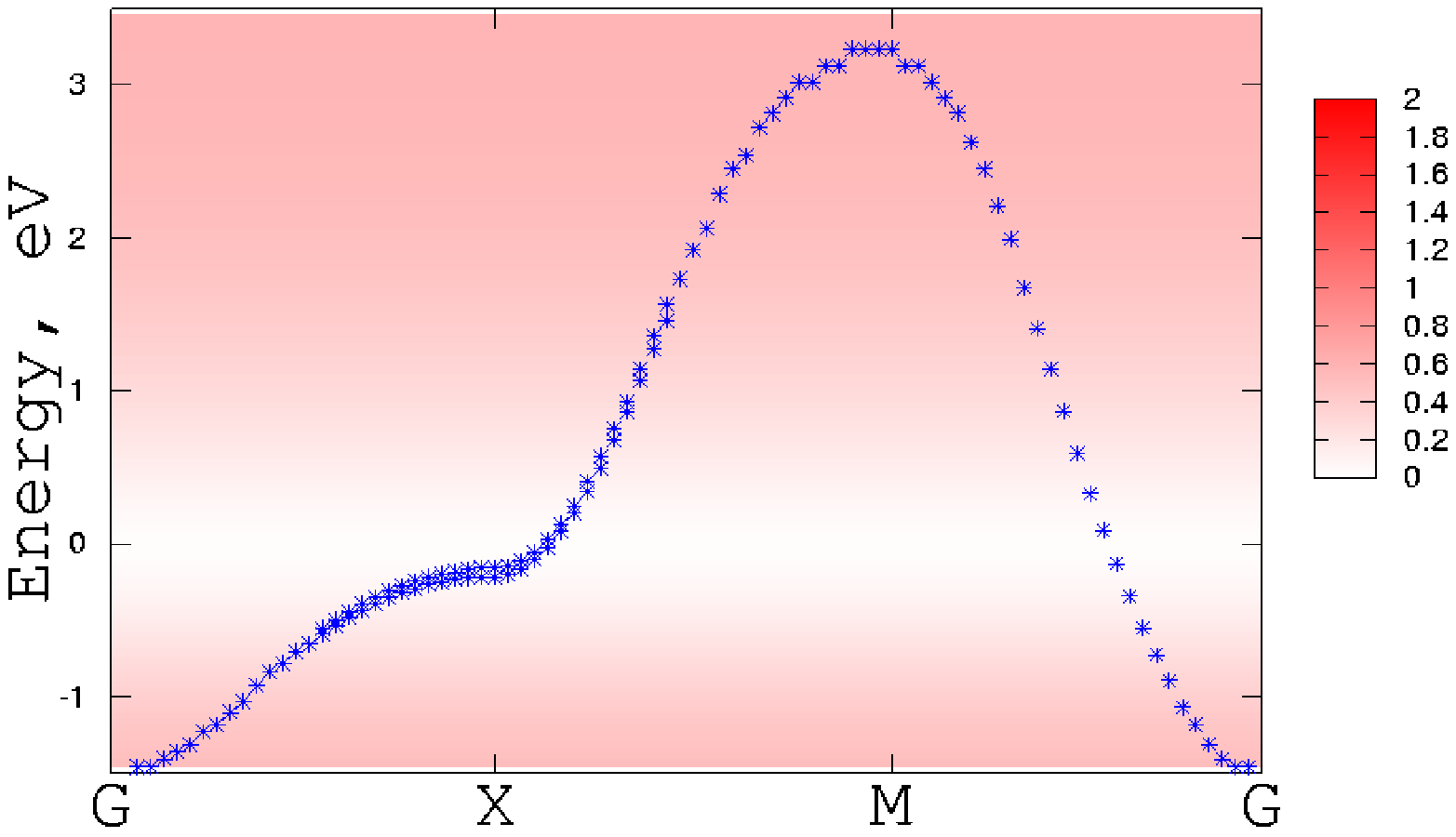}
\includegraphics[clip=true,width=0.4\textwidth]{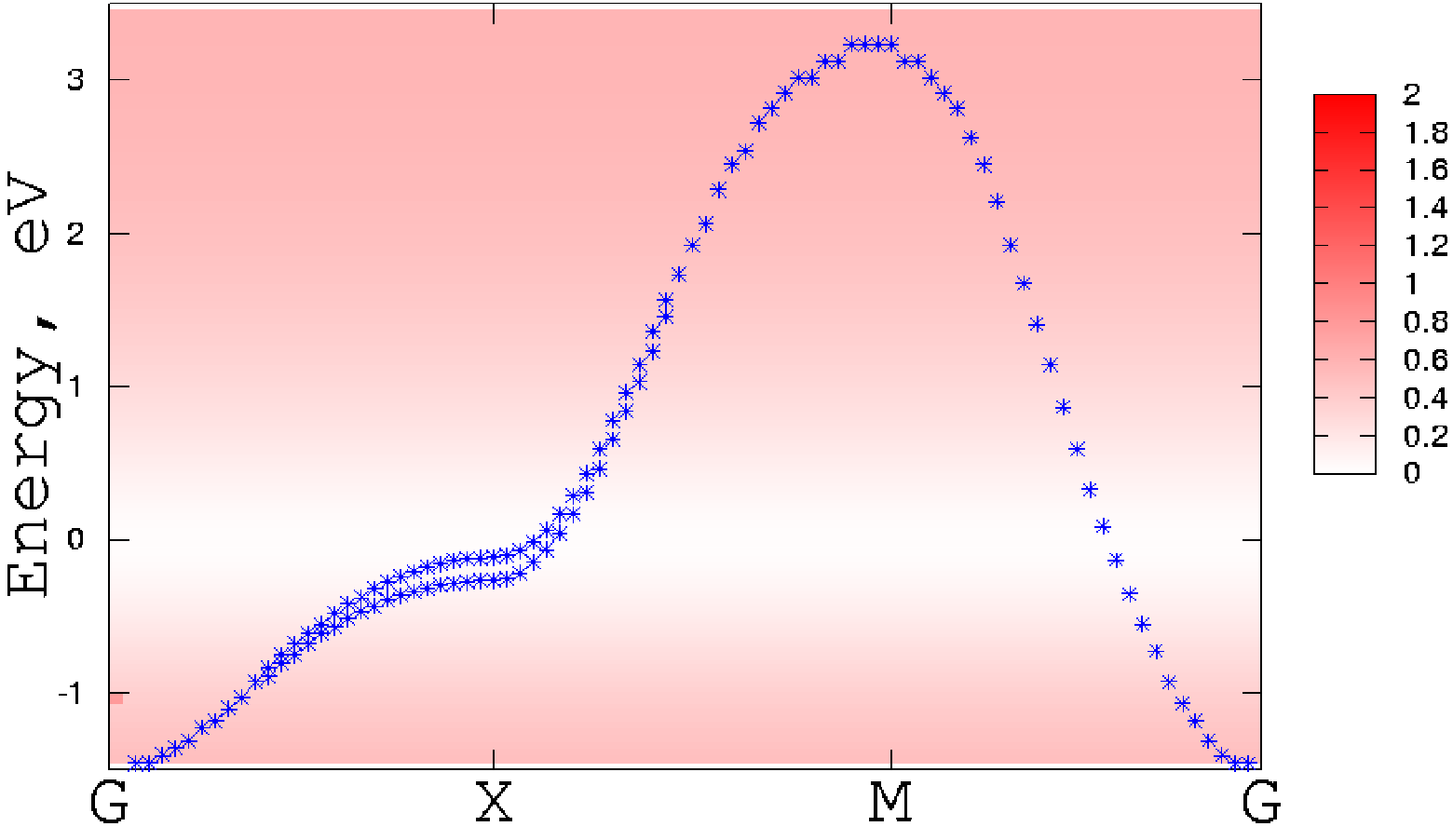}
\caption{LDA+DMFT quasiparticle bands for Bi2212 (crosses) along BZ high symmetry directions
for LDA-calculated value of $t_\perp$=0.03 eV (upper panel) and experimental value of $t^{exp}_\perp$=0.083 eV (lower panel)
(Coulomb interaction U=1.51 eV, filling n=0.85).
Zero of background (which is -1/$\pi$Im$\Sigma(\omega)$ - local DMFT
self-energy) corresponds to zero damping.}
\label{bands}
\end{figure}

\begin{figure}[htb]
\includegraphics[clip=true,width=0.4\textwidth]{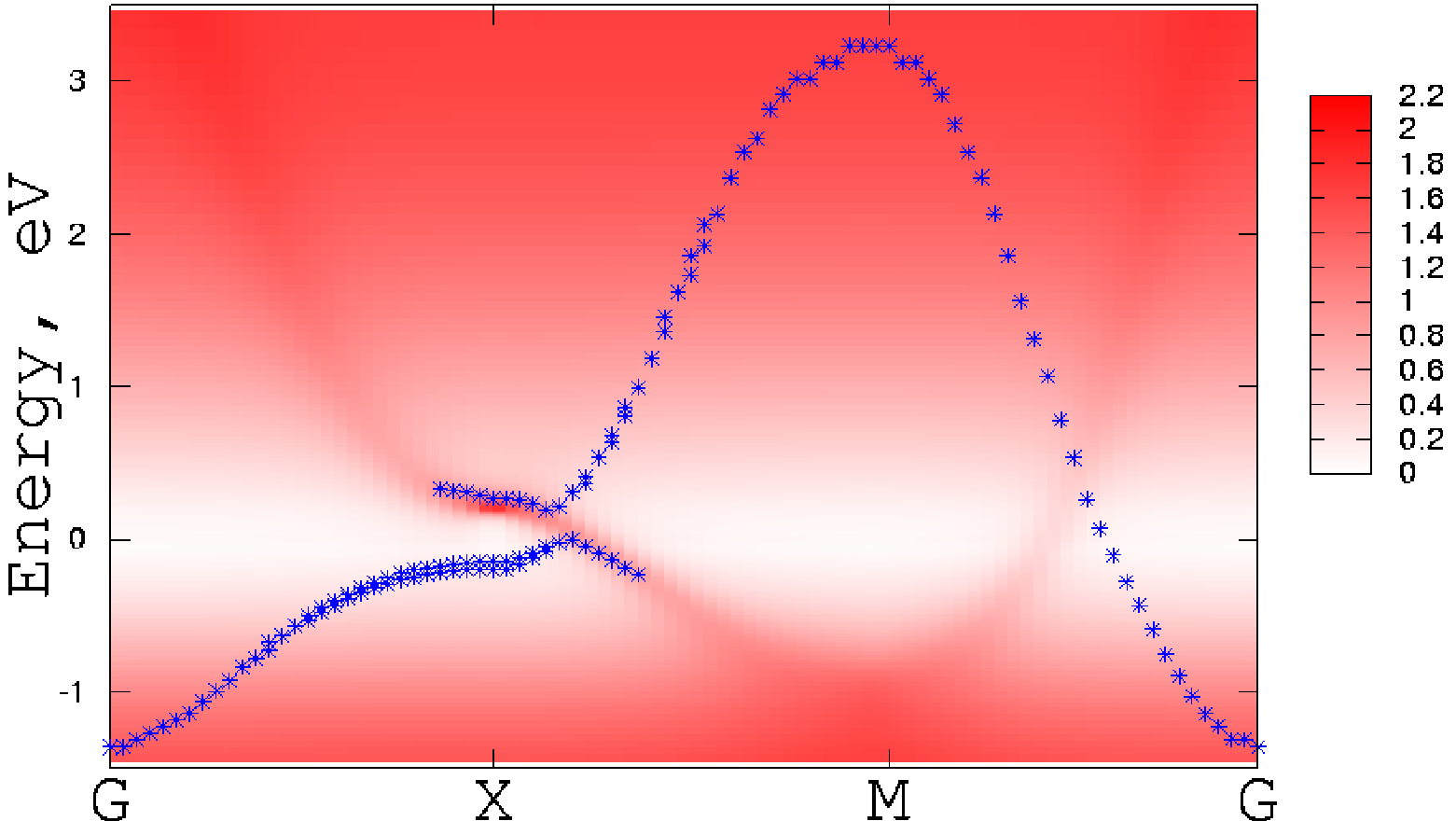}
\includegraphics[clip=true,width=0.4\textwidth]{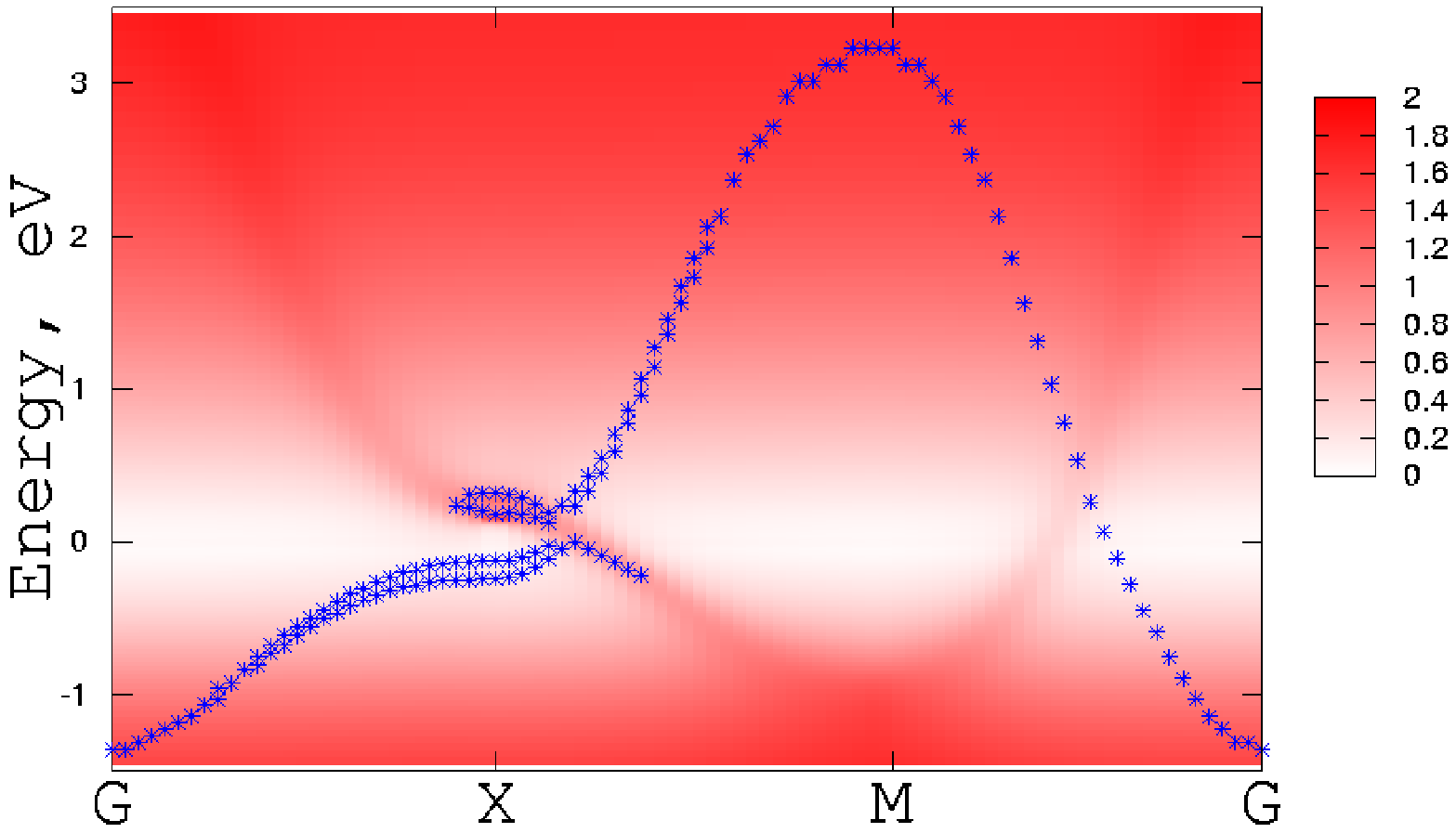}
\caption{LDA+DMFT+$\Sigma_{\bf k}$ quasiparticle bands for Bi2212 (crosses) along BZ high symmetry directions
for LDA-calculated value of $t_\perp$=0.03 eV (upper panel) and experimental value of $t^{exp}_\perp$=0.083 eV (lower panel)
(Coulomb interaction U=1.51 eV, filling n=0.85, pseudogap potential $\Delta$=0.21 eV, correlation length $\xi=5a$).
Zero of background (which is -1/$\pi$Im[$\Sigma(\omega)+\Sigma_{\bf k}(\omega)$] - additive
local and ``pseudogap'' self-energies) corresponds to zero damping.}
\label{sk_bands}
\end{figure}

In Figs.~\ref{bands} and~\ref{sk_bands} we present LDA+DMFT and 
LDA+DMFT+$\Sigma_{\bf k}$ quasiparticle bands dispersions (crosses) 
for Bi2212 effective $x^2-y^2$ orbital of Cu-3d shell
along the symmetry lines in the Brillouin zone (BZ) for $t_\perp$ and $t^{exp}_\perp$.
Background shows quasiparticle damping given by the imaginary part of 
additive $\Sigma(\omega)+\Sigma_{\bf k}(\omega)$ 
self-energy. The more intensive shade corresponds to the larger damping.  
In case of standard LDA+DMFT computations, neglecting non-local corrections 
(Fig.~\ref{bands}), one can clearly see that the damping is 
uniform over all BZ. This is due to local nature of conventional DMFT.  
Quasiparticles are well defined in narrow light region around zero energy 
(Fermi level).

When we introduce spatial inhomogeneity into DMFT bath within the 
LDA+DMFT+$\Sigma_{\bf k}$ approach the damping appears to be much stronger and
consequently non-uniform as seen in Fig.~\ref{sk_bands}.
Again quasiparticles are well defined close to the Fermi level.
But now the contour plot of 
Im$[\Sigma(\omega)+\Sigma_{\bf k}(\omega)]$ 
self-energy (damping) clearly shows so-called ``shadow band'' which looks 
like the quasiparticle band mirrored around the zero energy.  
In Fig.~\ref{sk_bands} we can also see pseudogap formation around X point.  
In our case shadow band is formed due to short-ranged AFM fluctuations.  
Close to X point BS effects are most pronounced.  
One can see that maxima of $A(\omega,{\bf k})$ belonging to the ``shadow band'' 
region are conserved only rather close to the X point, further away 
these maxima vanish due to large damping.  In the middle of MG direction 
we observe preformation of AFM insulating gap in the cross point of quasiparticle and
and ``shadow'' bands.

\subsection{Bi2212 LDA+DMFT+$\Sigma_{\bf k}$ spectral functions}
 
\begin{figure}[htb]
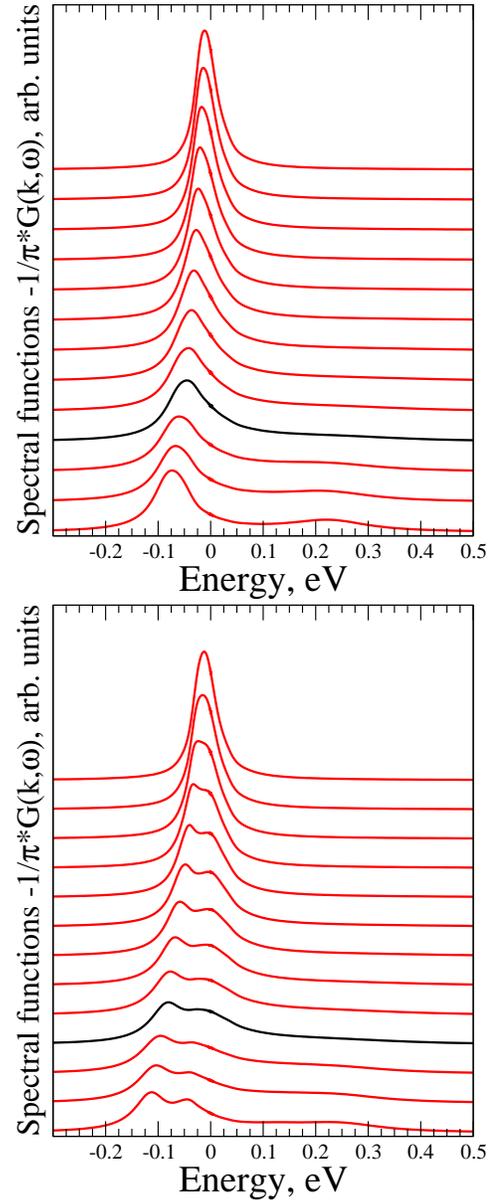

\includegraphics[clip=true,width=0.35\textwidth]{Bi2212_sd_pg033_x_02_bs003.eps}
\includegraphics[clip=true,width=0.35\textwidth]{Bi2212_sd_pg033_x_02_bs083.eps}
\caption{LDA+DMFT+$\Sigma_{\bf k}$ spectral densities for Bi2212 along of noninteracting FS in 1/8 of BZ
for LDA-calculated value of $t_\perp$=0.03 eV (upper panel) and experimental value of $t^{exp}_\perp$=0.083 eV (lower panel)
(Coulomb interaction U=1.51 eV, filling n=0.85, pseudogap potential $\Delta$=0.21 eV, correlation length $\xi=5a$).}
\label{sk_sd}
\end{figure}

To plot spectral functions $A(\omega,{\bf k})$ (\ref{specf}) we choose ${\bf k}$-points along the
1/8-th part of the ``bare'' Fermi surface within the first quadrant of the
Brillouin zone for given lattice spectra and filling. In Fig. \ref{sk_sd}
corresponding spectral functions for different strength
of bilayer splitting are shown. 

Close to the to nodal point (upper curve) spectral function in Fig. \ref{sk_sd}
has the typical Fermi--liquid behaviour, consisting of a
rather sharp peak close to the Fermi level. 
Going to the antinodal point (lower curve)
fluctuations becomes stronger and shift the sharp peak out of the Fermi level
down in energy. Simultaneously with the growth of fluctuation strength damping also grows,
so the peak becomes less intensive and more broad.  In the vicinity of the 
``hot--spot'' (black line) the shape of $A(\omega,{\bf k})$ is completely 
modified. Now $A(\omega,{\bf k})$ becomes double-peaked and 
non--Fermi--liquid--like.  Directly at the ``hot spot'', $A(\omega,{\bf k})$ 
has two peaks (second one is much less intensive)
situated symmetrically around the Fermi level and splitted from 
each other by $\sim 1.5\Delta$ \cite{Sch,KS}.

For the case of $t^{exp}_\perp$ (right panel of Fig. \ref{sk_sd})
behaviour is similar to one for  $t_\perp$ (upper panel of Fig. \ref{sk_sd}).
However now bilayer splitting strength is big enough to be resolved.
So the peak-dip-hump structure \cite{Bi2212} is formed on the edges of 
pseudogap.

\subsection{Bi2212 LDA+DMFT+$\Sigma_{\bf k}$ ARPES spectra}
 
\begin{figure}[htb]
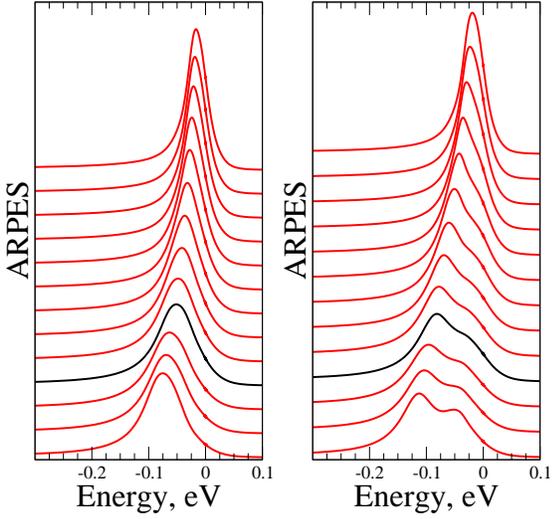

\includegraphics[clip=true,width=0.2\textwidth]{Bi2212_arpes_pg033_x_02_bs003.eps}
\includegraphics[clip=true,width=0.2\textwidth]{Bi2212_arpes_pg033_x_02_bs083.eps}
\caption{LDA+DMFT+$\Sigma_{\bf k}$ ARPES spectra for Bi2212 along of noninteracting FS in 1/8 of BZ
for LDA-calculated value of $t_\perp$=0.03 eV (left) and experimental value of $t^{exp}_\perp$=0.083 eV (right).
(Coulomb interaction U=1.51 eV, filling n=0.85, pseudogap potential $\Delta$=0.21 eV, correlation length $\xi=5a$).
Corresponding spectral function $A(\omega,{\bf k})$ is multiplied with Fermi
function at T$\sim$255K (the temperature of NRG calculations).}
\label{sk_arpes}
\end{figure}

Knowing $A(\omega,{\bf k})$ (\ref{specf}) we are now of course in a position 
to calculate angle resolved photoemission (ARPES) spectra, which are the most 
direct experimental way to observe pseudogap in real compounds. For that 
purpose, we only need to multiply our results for the spectral functions with 
Fermi function at temperature 255~K. The resulting 
LDA+DMFT+$\Sigma_{\bf k}$ ARPES spectra are presented in 
Fig.~\ref{sk_arpes}.  Again these spectra are drawn along 1/8 of 
non-interacting FS from antinodal (lower curve) to nodal point (upper curve). 
At the antinodal point we find well defined (sharp) quasiparticle peak close 
to the Fermi level. Moving towards the antinodal point the damping 
(widening) of this quasiparticle peak and its shift to higher
binding energies are observed. Such behaviour 
is typically obtained experimentally \cite{Bi2212}. 
To describe peak-dip-hump splitting 
resolved in experiment \cite{Bi2212} we take $t^{exp}_\perp$=0.083~eV 
\cite{Kordyuk03}. Indeed for $t^{exp}_\perp$ we get pronounced 
peak-dip-hump structure similar to experimental one \cite{Bi2212}.
It is recognized that our LDA-calculated 
$t_\perp$ is several times smaller and can not provide 
adequate description of the peak-dip-hump structure for ARPES data.
Notice that the intensity of antibonding branch is higher 
than for bonding one. It is opposite in the experiment.  We attribute this 
difference to the matrix elements effects which are not taken into account in 
the present work.

\begin{figure}[htb]
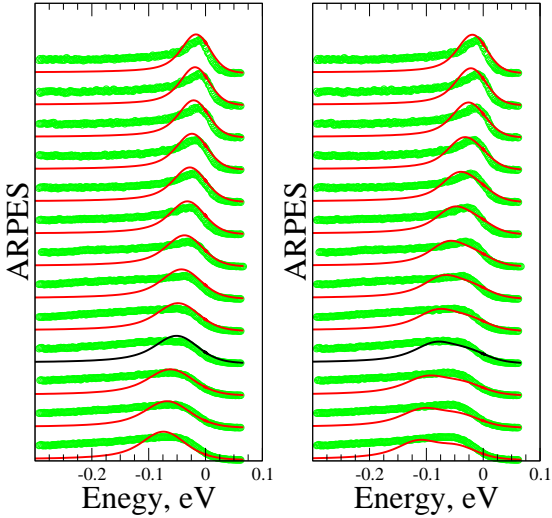

\includegraphics[clip=true,width=0.2\textwidth]{Bi2212_kaminski_arpes_BS003.eps}
\includegraphics[clip=true,width=0.2\textwidth]{Bi2212_kaminski_arpes_BS083.eps}
\caption{Comparison of LDA+DMFT+$\Sigma_{\bf k}$ ARPES spectra (solid lines) for Bi2212 along of noninteracting FS in 1/8 of BZ
for LDA-calculated value of $t_\perp$=0.03 eV (left) and experimental value of $t^{exp}_\perp$=0.083 eV (right)
with experimental ARPES (Ref.~\protect\cite{Kam2}) (circles).
(Coulomb interaction U=1.51 eV, filling n=0.85, pseudogap potential $\Delta$=0.21 eV, correlation length $\xi=5a$).
Corresponding spectral function $A(\omega,{\bf k})$ is multiplied by Fermi
function at T=140K (the temperature of experiment) and broadened with Gaussian to simulate experimental resolution
of 16 meV (Ref.~\protect\cite{Kam2}).}
\label{sk_exp}
\end{figure}

In Fig.~\ref{sk_exp} we show comparison of LDA+DMFT+$\Sigma_{\bf k}$ ARPES spectra
and experimental one from Ref.~\cite{Kam2} for Bi2212 measured along the Fermi surface.
Here spectral functions displayed in Fig.~\ref{sk_sd} are multiplied
with the Fermi function at experimental temperature T=140 K and convoluted
with Gaussian to simulate experimental resolution of 16 meV.~\cite{Kam2}
All theoretical ARPES curves after multiplication and broadening are normalized to 1.
Left and right panels of Fig.~\ref{sk_exp} correspond to the theoretical data
for $t_\perp$ and $t^{exp}_\perp$ values.
Both figures demonstrate semiquantitative agreement of our
theoretical results with the experiment.
Common trend for both panels is the damping of quasiparticle peak and its 
retreat to higher binding energies as we move from nodal to antinodal region.
Displacements of theoretical and experimental peaks on the left
panel of Fig.~\ref{sk_exp} are in good quantitative agreement.
However theoretical peaks are always a little bit sharper and narrower.
Take notice that the left panel demonstrates no BS effects.
On the right panel of Fig.~\ref{sk_exp}
we found slightly better agreement of intensities due to
bigger BS value $t^{exp}_\perp$. But for the ${\bf k}$-values between
the ``hot spot'' and the antinodal point there we have some lack of
spectral weight close to the Fermi level. Also for these
${\bf k}$-values we observe some reminiscence of the bilayer splitting.
After all one can infer that BS effects do not change the line shape of
our ARPES spectra significantly and for both cases we obtain rather 
satisfactory agreement with the experiment.

\subsection{Bi2212 LDA+DMFT+$\Sigma_{\bf k}$ Fermi surface}

In the following we characterize renormalized Fermi surfaces (FS) by intensity plots 
of spectral density at zero frequency $A(\omega=0,{\bf k})$ (which for the 
free-electron case just follow the ``bare'' Fermi surface).

\begin{figure}[htb]
\includegraphics[clip=true,width=0.45\textwidth]{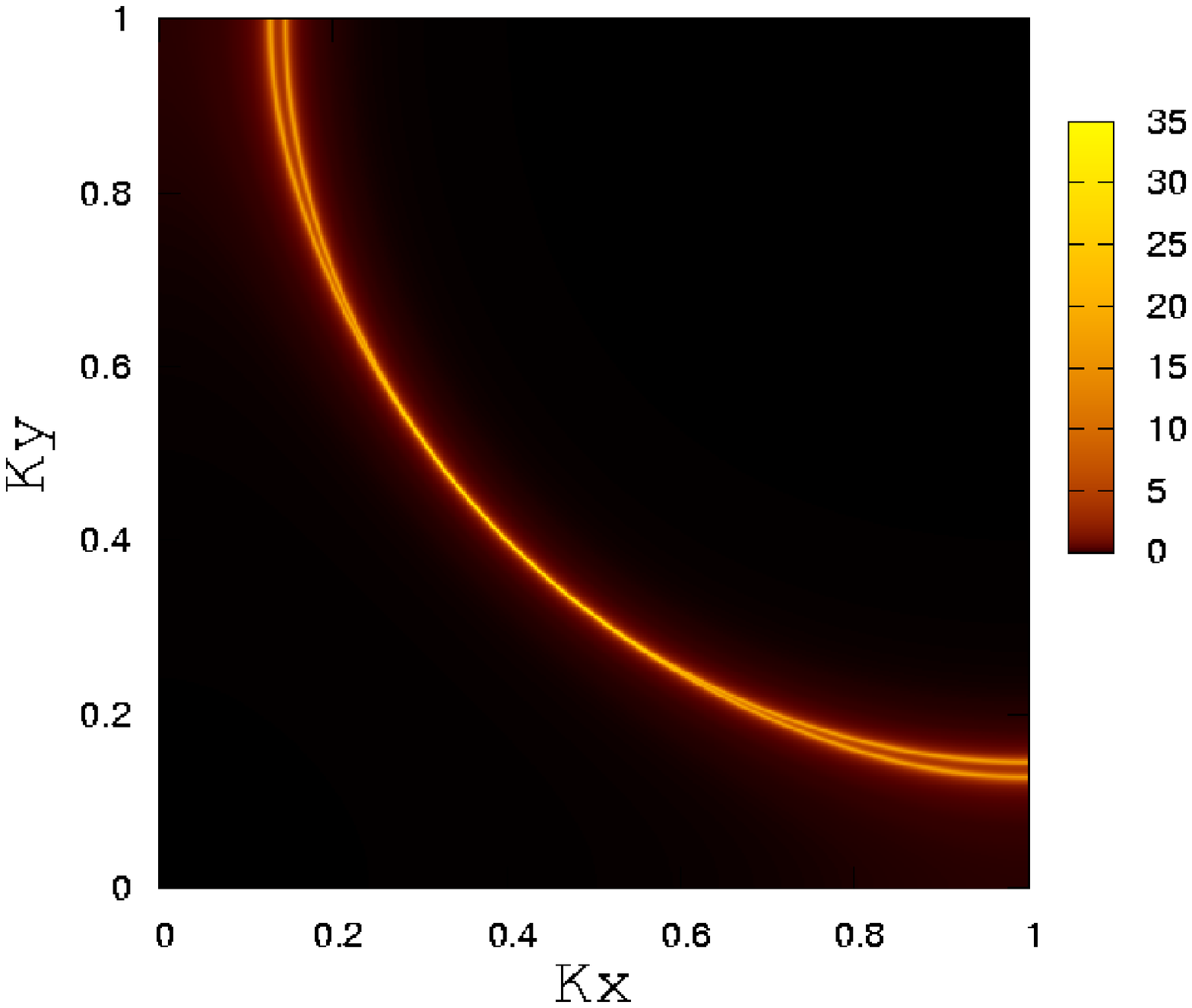}
\includegraphics[clip=true,width=0.45\textwidth]{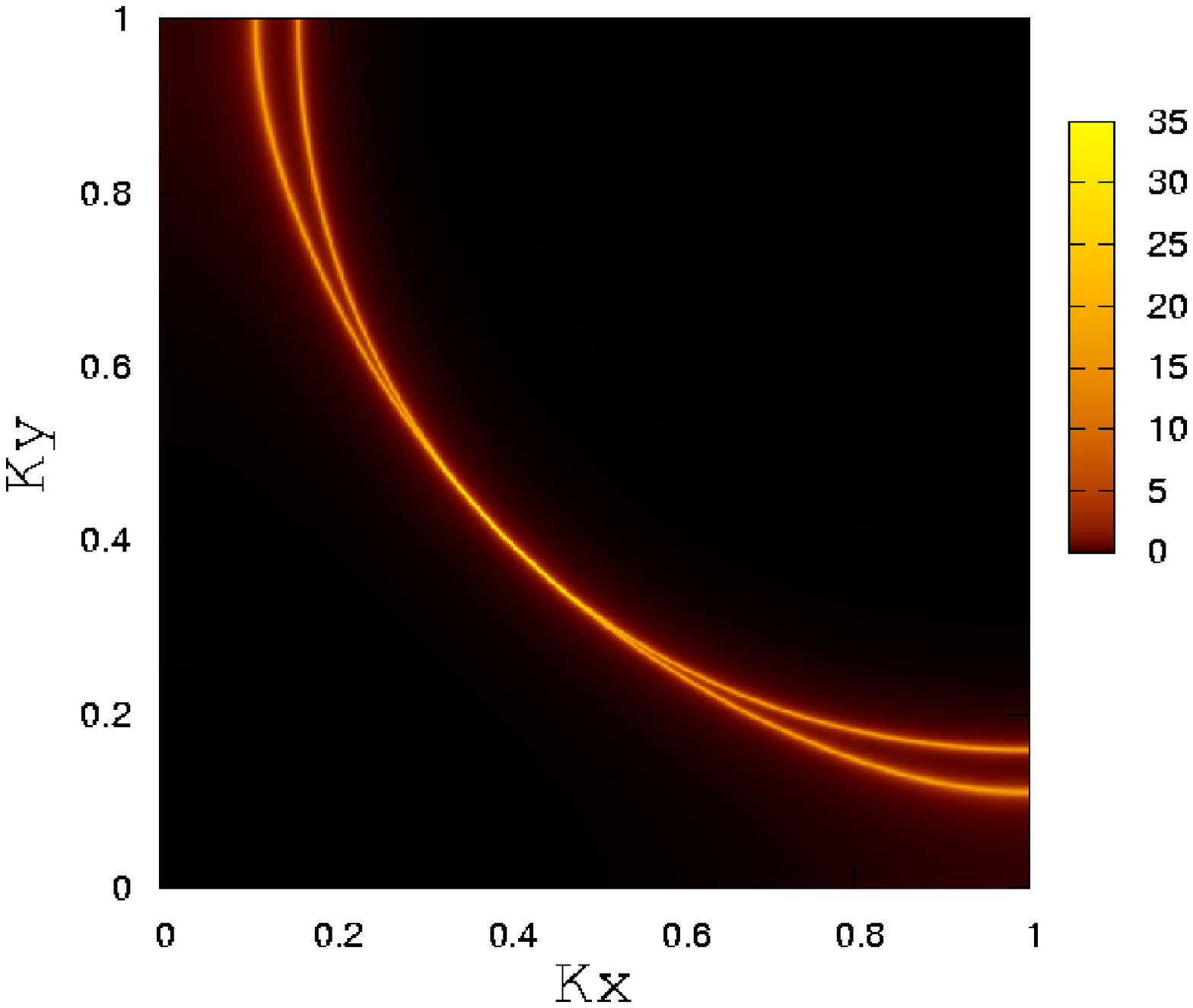}
\caption{LDA+DMFT Fermi surfaces for Bi2212 within 1/4 of BZ
($k_x,k_y$ in units of $\pi/a$)
for LDA-calculated value of $t_\perp$=0.03 eV (upper panel) and experimental value of $t^{exp}_\perp$=0.083 eV (lower panel)
(Coulomb interaction U=1.51 eV, filling n=0.85).}
\label{fs}
\end{figure}

\begin{figure}[htb]
\includegraphics[clip=true,width=0.45\textwidth]{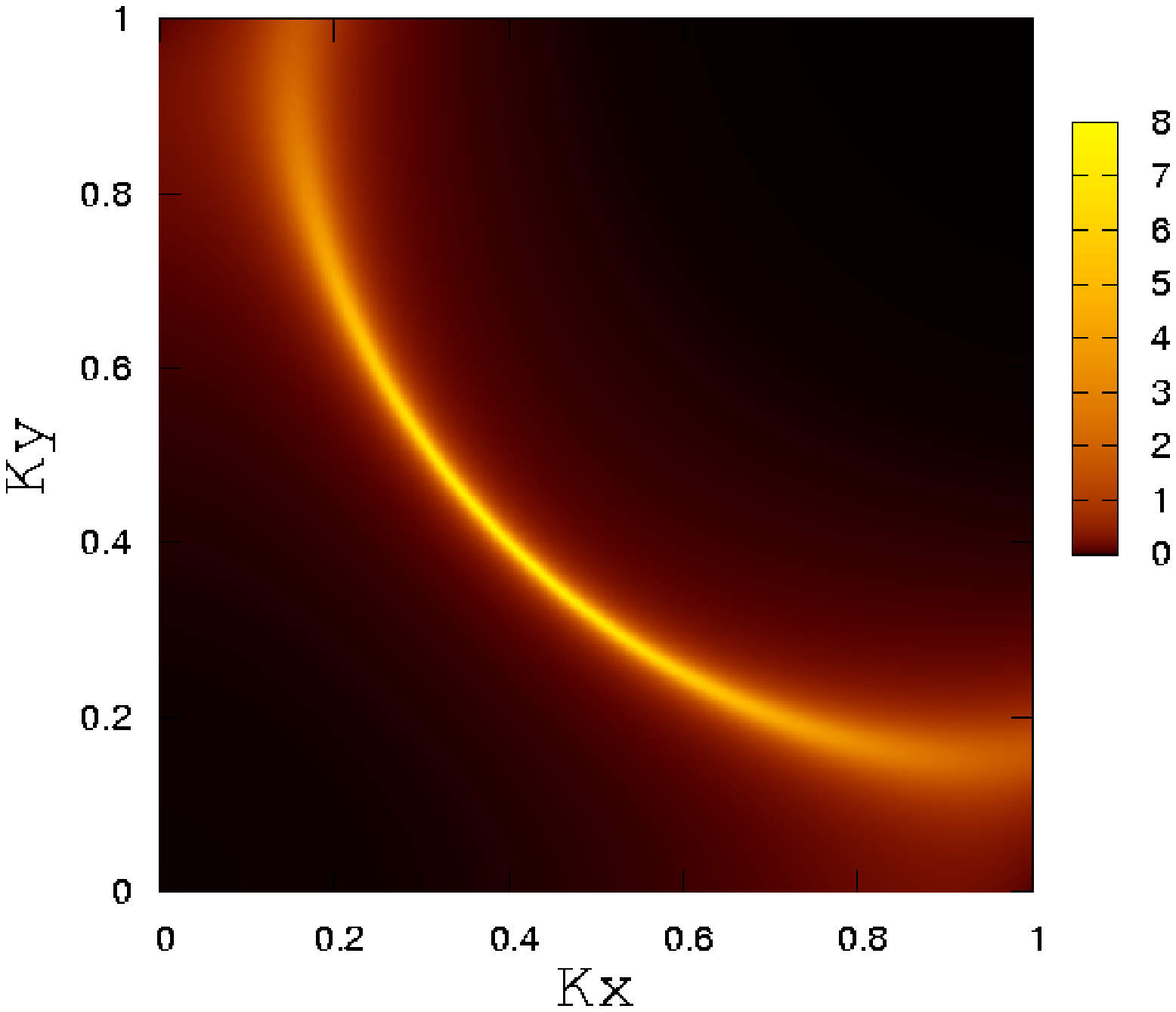}
\includegraphics[clip=true,width=0.45\textwidth]{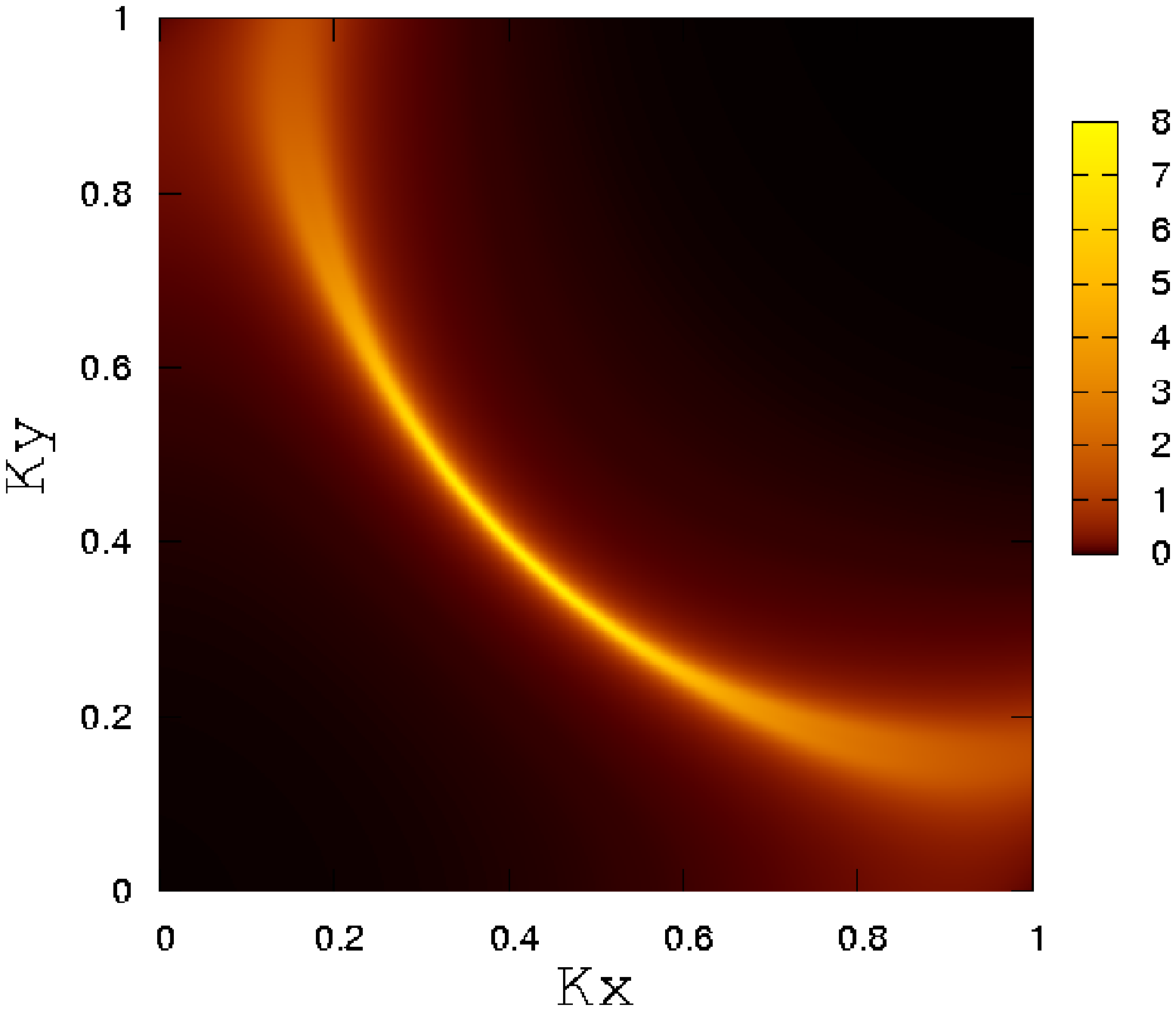}
\caption{LDA+DMFT+$\Sigma_{\bf k}$ Fermi surfaces for 
Bi2212 within 1/4 of BZ ($k_x,k_y$ in inits of $\pi/a$)
for LDA-calculated value of $t_\perp$=0.03 eV (upper panel) and experimental value of $t^{exp}_\perp$=0.083 eV (lower panel)
(Coulomb interaction U=1.51 eV, filling n=0.85, pseudogap potential $\Delta$=0.21 eV, correlation length $\xi=5a$).}
\label{sk_fs}
\end{figure}

In the Figs.~\ref{fs} and~\ref{sk_fs} we display thus defined LDA+DMFT and 
LDA+DMFT+$\Sigma_{\bf k}$ Fermi surfaces for Bi2212. 
LDA+DMFT FS has the LDA shape, as it should be 
within DMFT (see Fig.~\ref{fs}). Slight broadening close to the borders of BZ is 
because of BS effects. Non-zero width of FS (in contrast to LDA) comes from 
finite damping due to interaction and temperature. 
For LDA+DMFT+$\Sigma_{\bf k}$ FS (see Fig.~\ref{sk_fs}) one can see significant 
``destruction'' effects in the vicinity of the antinodal point induced by pseudogap 
fluctuations.  From comparison of upper and lower panels of Fig.~\ref{sk_fs} 
one can conclude that for strongly correlated case BS effects alone are not 
enough to describe experimentally observed FS ``destruction'' close to the 
borders of BZ and formation of ``Fermi arcs'' around the nodal point,
as observed in ARPES experiments \cite{Bi2212}.  
We found that FS shape is rather insensitive to BS 
strength since pseudogap fluctuations are much stronger than 
bilayer splitting and hide it.
Though BS at the BZ boundaries slightly amplifies pseudogap effects.
Thus the account of pseudogap (AFM) 
fluctuations seems to be necessary to describe experimental picture.

\subsection{Bi2212 LDA+DMFT+$\Sigma_{\bf k}$ anisotropy of static scattering}

Strong anisotropy of (quasi) static scattering was observed in Bi2212 system
in ARPES experiments in Refs. \cite{Valla, Kam2, Kam1} and attributed to
scattering by planar impurities \cite{VA1, VA2}.
Here we show that this effect can be naturally explained by (quasi) static
scattering by pseudogap fluctuations.

\begin{figure}[htb]
\includegraphics[clip=true,width=0.4\textwidth]{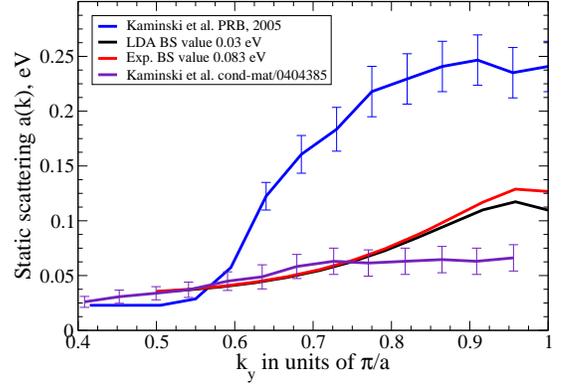}
\caption{Comparison of experimental and theoretical
LDA+DMFT+$\Sigma_{\bf k}$ static scattering $a({\bf k})=-1/\pi {\rm Im}
\Sigma(0)+\Sigma_{\bf k}(0)$
for Bi2212 within 1/8 of BZ
for LDA-calculated value of $t_\perp$=0.03 eV (light curve) and experimental value of $t^{exp}_\perp$=0.083 eV (dark curve)
(Coulomb interaction U=1.51 eV, filling n=0.85, pseudogap potential $\Delta$=0.21 eV, correlation length $\xi=5a$).}
\label{sk_ak}
\end{figure}

Our LDA+DMFT+$\Sigma_{\bf k}$ calculated (quasi) static scattering defined as
$a({\bf k})=\Sigma(0)+\Sigma_{\bf k}(0)$
is plotted in Fig.~\ref{sk_ak}, together with experimental data of Refs.
\cite{Kam2, Kam1}. Here {\bf k}-points are taken along 1/8 of 
non-interacting FS. We detect our results to mediate
experimental data of Refs. \cite{Kam2,Kam1}, while the difference 
between latter remains itself unexplained. 

In our opinion,  anisotropy of (quasi) static scattering $a({\bf k})$ naturally follows 
from anisotropic renormalization of electronic spectrum due to pseudogap 
fluctuations, which directly follows from our ``hot spot'' like model
\cite{Sch,KS}.

Despite overall behaviour is analogous to one obtained in the experiment
there is a need for further studies of possible relevance of matrix
elements effects in ARPES, as well as that of additional scattering by random
static impurities \cite{FNT}.

\section{Conclusion}
\label{concl}

Present investigation is aimed to describe pseudogap regime of  high-T$_c$ cuprate 
Bi$_2$Sr$_2$CaCu$_2$O$_{8-\delta}$ (Bi2212) from first principles.
For this purpose we employ novel generalized $ab~initio$ 
LDA+DMFT+$\Sigma_{\bf k}$ hybrid scheme.
This scheme based on the strategy of most powerfull computational tool for real 
correlated materials: local density approximation (LDA) + dynamical mean-field 
theory (DMFT).  Here we supply conventional LDA+DMFT equations with an additional 
(momentum dependent) self-energy $\Sigma_{\bf k}$ in the spirit of our 
recently proposed DMFT+$\Sigma_{\bf k}$ approach.  ``External'' self-energy 
$\Sigma_{\bf k}$ is chosen here to describe non-local dynamical correlations 
induced by short-ranged collective Heisenberg-like antiferromagnetic spin 
fluctuations (in static Gaussian approximation of Refs. \cite{Sch,KS}).  
Necessary Bi2212 material specific model parameters for effective $x^2-y^2$
orbital of Cu-3d shell, e.g. the 
values of intra- and interlayer hopping integrals, local 
Coulomb interaction U and pseudogap potential $\Delta$ 
were calculated within LDA and LDA+DMFT. On the basis of 
LDA+DMFT+$\Sigma_{\bf k}$ computations we obtain densities of states, spectral functions 
$A(\omega,{\bf k})$ which allow one to visualize quasiparticle bands 
dispersion and damping, Fermi surface (FS), anisotropy of static scattering $a({\bf k})$
and ARPES spectra accounting for pseudogap and bilayer splitting effects for 
normal (slightly) underdoped Bi2212 ($\delta$=0.15).  
It is found that on the DOS level BS and pseudogap effects are separated in energy
and hardly affect each other. We showed that LDA+DMFT+$\Sigma_{\bf k}$ 
describes strong scattering at Brillouin zone boundaries as pure 
manybody effect.  LDA+DMFT+$\Sigma_{\bf k}$ Fermi surface in presence of 
the pseudogap fluctuations is almost insensitive to the BS 
strength. Thus the BS effects alone are not enough to describe the Fermi surface destruction
(though amplifies it)
and additional source of electron scattering is required (for example, AFM short-range fluctuations).
The only place where BS effects play significant role is
formation of the experimentally observed peak-dip-hump structure
in ARPES spectra.
To this end the LDA-calculated value of bilayer splitting is found to be rather small to 
describe this effect.  Results obtained 
are in good semiqualitative agreement with various recent ARPES experiments.

At present there are several alternative points of view on the possible 
explanation of Fermi surface destruction, formation of shadow Fermi bands etc.
Recently the analysis of the effect of three-dimensionality on the ARPES spectra 
was presented for  Bi2212 in Ref. \cite{bansil_05}. It was shown that in a 
quasi-2D system, the weak $k_z$-dispersion can lead to Fermi surface maps 
similar to those observed in the experiment. 
This FS broadening mechanism does not have the manybody origin.
The authors of Ref. \cite{mans_06} have shown that shadow Fermi surface 
in Bi2212 can be interpreted as an intrinsic feature of the initial electronic 
spectrum arising from bulk, orthorhombic distortions located primarily in the 
BiO planes, but most definitely felt throughout the three-dimensional crystal.
All these effects are not considered here thus remain for further investigations.
Apparently, in a real
system these mechanisms combine with those described above leading to a 
complete picture of electronic structure of Bi2212.

\section{Acknowledgements}
We thank Thomas Pruschke for providing us the NRG code.
This work is supported by RFBR grants 05-02-16301, 05-02-17244, 06-02-90537, RAS programs 
``Quantum macrophysics'' and ``Strongly correlated electrons in 
semiconductors, metals, superconductors and magnetic materials'', Dynasty 
Foundation, Grant of President of Russia MK.2118.2005.02, interdisciplinary 
UB-SB RAS project, Russian Science Support Foundation.

\newpage

\newpage

\newpage

\newpage

\newpage

\newpage

\newpage

\newpage

\newpage

\begin{table}
\label{param}
\caption {Calculated energetic model parameters for Bi2212 (eV). First four Cu-Cu inplain hopping integrals
$t$, $t^{\prime}$, $t^{\prime\prime}$, $t^{\prime\prime\prime}$, interplain hopping value~$t_\perp$,
local Coulomb interaction~$U$ and pseudogap potential~$\Delta$.}
\begin{tabular}{| c | c | c | c | c | c | c |}
\hline
$t$  & $t^{\prime}$  & $t^{\prime\prime}$ &  $t^{\prime\prime\prime}$ &  $t_\perp$ & $U$ & $\Delta$\\
\hline
-0.627 & 0.133 & 0.061 & -0.015 & 0.03 & 1.51 & 0.21\\
\hline
\end{tabular}
\end{table}

\end{document}